\documentclass[aps,prb,twocolumn,showpacs,preprintnumbers,amsmath,amssymb,longbibliography]{revtex4-2}

\newcommand{\expect}[1]{\langle #1 \rangle}

\newcommand{\be}{\begin{equation}}
\newcommand{\ee}{\end{equation}}
\newcommand{\bea}{\begin{eqnarray}}
\newcommand{\eea}{\end{eqnarray}}
\newcommand{\eq}[1]{Eq.~(\ref{#1})}
\newcommand{\fig}[1]{Fig.~\ref{#1}}
\newcommand{\figs}[1]{Figs.~\ref{#1}}
                             
\newcommand{\e}{\varepsilon}
\newcommand{\w}{\omega}
\newcommand{\s}{\sigma}
\newcommand{\G}{\Gamma}
\newcommand{\up}{\uparrow}
\newcommand{\down}{\downarrow}

\newcommand{\new}[1]{#1}

\newcommand{\Sn}{S_{\rm n}}
\newcommand{\Sd}{S_{\rm d}}

  \newcommand{\Sec}[1]{Sec.~\ref{#1}}
  
  \newcommand{\Eq}[1]{Eq.~\eqref{#1}}

  \newcommand{\Fig}[1]{Fig.~\ref{#1}}

\usepackage{float}
\usepackage[dvipsnames]{xcolor}
\usepackage{ulem}
\usepackage{verbatim} 
\usepackage{amsmath}
\usepackage{graphicx}
\usepackage{dcolumn}
\usepackage{bm}
\usepackage{hyperref}

\definecolor{oucrimsonred}{rgb}{0.6, 0.0, 0.0}

\begin{document}

\title{Nonequilibrium Seebeck effect \new{and thermoelectric efficiency} \\of Kondo-correlated molecular junctions}

\author{Anand Manaparambil}
\email{anaman@amu.edu.pl}
\affiliation{Institute of Spintronics and Quantum Information,
Faculty of Physics, 
Adam Mickiewicz University,
Uniwersytetu Pozna\'nskiego 2, 61-614 Pozna\'n, Poland}

\author{Ireneusz Weymann}
\affiliation{Institute of Spintronics and Quantum Information,
Faculty of Physics, 
Adam Mickiewicz University,
Uniwersytetu Pozna\'nskiego 2, 61-614 Pozna\'n, Poland}

\date{\today}


\begin{abstract}
We theoretically study the nonequilibrium thermoelectric transport properties of 
a strongly-correlated molecule (or quantum dot) embedded in a tunnel junction.
Assuming that the coupling of the molecule to the contacts is asymmetric,
we determine the nonlinear current driven by the voltage and temperature
gradients by using the perturbation theory. However,
the subsystem consisting of the molecule strongly
coupled to one of the contacts is solved by using the numerical renormalization
group method, which allows for accurate description 
of Kondo correlations. We study the temperature gradient and voltage dependence 
of the nonlinear and differential Seebeck coefficients for
various initial configurations of the system.
In particular, we show that in the Coulomb blockade regime with singly occupied molecule,
both thermopowers exhibit sign changes  due to the Kondo correlations
at nonequilibrium conditions.
\new{Moreover, we determine the nonlinear heat current and thermoelectric efficiency,
	demonstrating that the system can work as a heat engine
	with considerable efficiency, depending on the transport regime.}
\end{abstract}

\maketitle

\section{\label{sec:level1}Introduction}

Thermoelectric properties of nanoscale systems
have become a subject of extensive studies \cite{Dhar2008Sep,Dubi2009Jan,Dubi2011Mar,Benenti2017Jun}.
This is because such structures, due to their reduced dimensions,
allow for obtaining thermal response much exceeding
that obtained in bulk materials \cite{MahanSofo}.
In this regard, a special role is played by zero-dimensional systems,
such as molecules or quantum dots,
in which discrete energy spectrum is relevant for
efficient energy filtering and obtaining considerable figure of merit.
Thermopower of such systems has already been 
explored theoretically \cite{Krawiec2006May, Franco2008Jul, Liu2010Jun,Costi2010Jun, Nguyen2010Sep,  Eckern2020Jan},
both in the weak and strong coupling regimes,
as well as experimentally \cite{Svilans2016Dec,Svilans2018Nov,Dutta2019Jan}.
A special attention has been paid to strong electron correlation regime,
where the Kondo effect can emerge at sufficiently low temperatures \cite{Kondo1964Jul,Hewson_1993}.
This is due to the fact that the analysis
of the temperature dependence of the Seebeck effect
can provide additional information about
the Kondo correlations in the system \cite{Costi2010Jun,Svilans2018Nov,Dutta2019Jan}.
In particular, sign changes of linear thermopower
as a function of temperature were shown to indicate
the onset of the Kondo correlations in the system \cite{Svilans2018Nov,Dutta2019Jan}.
From theoretical side, accurate description
of thermoelectric phenomena in the strong correlation regime
requires using sophisticated numerical methods,
therefore such considerations have been mostly limited to the linear response regime
\cite{Costi2010Jun,Weymann2013Aug,Wojcik2016Feb,
	Costi2019Oct,Costi2019Oct2,Manaparambil2021Apr},
while much less attention has been paid to the far-from-equilibrium regime \cite{Talbo2017Nov,Eckern2020Jan}.

The goal of this work is therefore to shed more light on 
the nonequilibrium thermoelectric characteristics in systems where Kondo correlations are crucial.
For that, we consider a molecule (or a quantum dot) asymmetrically attached to external contacts.
Such system's geometry allows us to incorporate strong electron correlation effects
in far-from-equilibrium conditions in a very accurate manner.
The electronic correlations give rise to the development of the Kondo phenomenon,
which arises due to the strong coupling to one of the contacts, whereas
the second contact, serves as a weakly coupled probe. In such scenario,
we can make use of the numerical renormalization group (NRG)
method \cite{Wilson1975Oct, Bulla2008Apr, FlexibleDMNRG} to solve the strongly-coupled subsystem,
while the nonequilibrium current flowing through the whole system,
triggered by voltage and/or temperature gradients,
is evaluated based on the perturbation theory with respect to 
the weakly attached electrode.
We show that with this method we can determine the thermoelectric effects
at large and finite temperature and potential gradient without losing the Kondo correlations.
\new{
We also note that accurate quantitative calculation of
nonequilibrium thermoelectric transport in the case of {\it symmetrically-coupled} systems
poses a great challenge that could be addressed by recently developed hybrid approach involving 
time-dependent density matrix renormalization group and NRG \cite{Manaparambil2022Sep,Schwarz2018Sep}.
}

For the considered system here, we first study the voltage dependence of the differential conductance,
demonstrating suppression of the zero-bias Kondo peak with
increasing the temperature gradient.
We then focus on the analysis of nonequilibrium and differential Seebeck effects,
analyzing at the beginning the case of finite temperature gradient
within linear response in applied voltage. Furthermore,
we examine the behavior of the thermopower in the nonlinear voltage and temperature gradient regime,
predicting new sign changes associated with Kondo correlations.
\new{
	We also calculate the heat currents and the power generated by the device
	as well as the corresponding thermoelectric efficiency,
	revealing regimes of large efficiency depending on the transport regime.
}
Finally, assuming realistic junction parameters,
we consider the thermoelectric transport properties
including the voltage dependence of the molecule's orbital level.

The paper is organized as follows. In \Sec{sec:theory}, we discuss the model and method used
in the calculations along with the definitions for thermopower in out-of-equilibrium settings.
The main results and discussions are presented in \Sec{sec:results},
which begins with the analysis of the electronic transport
under a finite potential bias and temperature gradient.
The nonequilibrium thermoelectric coefficients are then studied,
first at zero-bias, and then generalized to the finite bias and temperature conditions.
\new{
We also determine the behavior of the nonequilibrium heat current and the
thermoelectric efficiency of the system under various parameter regimes.}
Finally, the paper is summarized in \Sec{sec:summary}.

\section{\label{sec:theory}Theoretical Framework}
\subsection{Model and Hamiltonian}

\begin{figure}[t]
	\centering
\includegraphics[width=0.85\columnwidth]{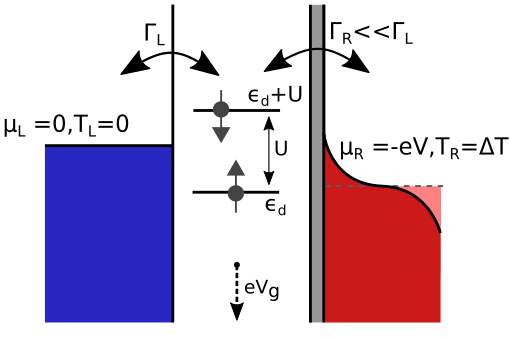}
\caption{The schematic of the considered system.
It consists of a molecule (or a quantum dot)
characterized by an orbital level of energy $\e_d$ and Coulombic repulsion $U$.
The molecule is strongly coupled to the left lead with coupling strength $\Gamma_L$,
while the coupling to the right contact $\Gamma_R$ is much weaker.
\new{$V_g$ stands for the gate voltage, which can be used to tune
the position of the orbital level. $\mu_L$ ($\mu_R$) and $T_L$ ($T_R$)
denote the chemical potential and temperature of the left (right) lead, respectively.}
The left contact is grounded and kept at a constant temperature $T_L=0$,
while the right contact is subject to voltage bias $V$ and temperature gradient $\Delta T$.
}
\label{Fig:E_schematic}
\end{figure}

The considered system consists of a molecule
(or a quantum dot) asymmetrically coupled to two metallic leads,
as schematically shown in  Fig.~\ref{Fig:E_schematic}. The molecule is described 
by an orbital level of energy $\e_d$ and Coulomb
correlations denoted by $U$.
\new{
This orbital level may correspond to the lowest unoccupied orbital level (LUMO level)
of the molecule.}
Transport through the system can be induced
and controlled by applying a finite bias voltage $V$
and/or a temperature gradient $\Delta T$ across the leads.
We assume that the left contact is grounded and kept at a constant temperature $T$,
while the right contact is subject to $V$ and $\Delta T$, see Fig.~\ref{Fig:E_schematic}.
\new{
The temperature $T$ is assumed to be much smaller than 
the characteristic energy scale of the Kondo effect, i.e. in practical
calculations we set $T\to 0$.
}
Moreover, it is assumed that the coupling to the left lead ($\Gamma_L$) is much stronger than
the coupling to the right electrode ($\Gamma_R$).
Such an asymmetry is frequently present in various molecular junctions \cite{Bauer2013Feb,Xu2016Apr,Gruber2018Sep,Zonda2021Jul,Xing2022Feb},
it can be also generated in artificial heterostructures comprising
e.g. a quantum dot \cite{Csonka2012May, PerezDaroca2018Apr, Tulewicz2021Jul}.
Furthermore, it can be also encountered in studies of adatoms with scanning tunneling spectroscopy.
Under this assumption, we determine the current flowing through the system using the perturbation theory in $\Gamma_R$,
while the subsystem consisting of molecule strongly coupled
to the left lead is treated exactly with the aid of the numerical renormalization group method
\cite{Wilson1975Oct, Bulla2008Apr, FlexibleDMNRG}.
\new{In other words, we include the lowest-order processes between the
molecule and the right electrode, while the tunneling processes between the molecule
and the left lead are taken into account in an exact manner.
}

The Hamiltonian of the molecule coupled to the left lead can be expressed as
\cite{Anderson1970Dec}
\begin{eqnarray}
\label{Eq:HL}
H_L &=& \varepsilon_d n + U n_\up n_\down
+\sum_{k\s} \e_{Lk\sigma} c^{\dagger}_{Lk \s} c_{Lk\s}\nonumber\\
&&+\sqrt{\frac{\Gamma_L}{\pi\rho_L}} \sum_{k\s} ( d^{\dagger}_{\s} c_{Lk\s} + c^{\dagger}_{Lk\s}d_{\s} ),
\end{eqnarray}
\new{
	where the first two terms model the orbital level with energy $\e_d$
and Coulomb correlations $U$},
with $n=n_\uparrow+n_\downarrow$, $n_\sigma = d_\sigma^\dagger d_\sigma$,
and $d_\sigma^\dagger$ ($d_\sigma$) being the creation (annihilation)
operator for spin-$\sigma$ electrons on the orbital level.
The creation (annihilation) operator for an electron of spin $\sigma$,
momentum $k$ and energy $\e_{\alpha k\sigma}$ in the lead $\alpha$ is
denoted by $c_{\alpha k \sigma}^\dagger$ ($c_{\alpha k \sigma}$).
\new{The third term in \eq{Eq:HL} describes the left lead in the free quasiparticle
approximation, while the last term models the tunneling processes between
the molecule and left lead.}
The density of states of the lead $\alpha$ is described by $\rho_\alpha$.
\new{Here we use the wide-band approximation under which the couplings are energy independent.}
On the other hand, the Hamiltonian of the right contact can be written as \cite{Haug},
\be
H_R = \sum_{k\s} \varepsilon_{R k\s} c^{\dagger}_{Rk\s} c_{Rk\s} - eV \sum_{k\s} c^{\dagger}_{Rk\s} c_{Rk\s},
\ee
where $V$ is the applied bias voltage and $e$ stands for the elementary charge.
Then, the tunneling processes between the left and right part of the system
can be described by the following tunneling Hamiltonian \cite{Haug}
\be
H_T = \sqrt{\frac{\Gamma_R}{\pi\rho_R}} \sum_{k\s} ( d^{\dagger}_{\s} c_{Rk\s} + c^{\dagger}_{Rk\s} d_{\s}).
\ee
Thus, the total Hamiltonian is given 
by a sum of three terms, the strongly coupled left part, $H_L$, the weakly coupled right lead, $H_R$,
and the term accounting for tunneling between both parts, $H_{T}$,
$H = H_L + H_R + H_{T}$. In what follows,
to determined the tremoelectric transport properties,
we perform a perturbation expansion in $H_{T}$.

\subsection{Nonequilibrium transport coefficients}

\subsubsection{Electric current}

The assumption of the weak coupling $\Gamma_R$ to the right subsystem allows us to perform
a perturbative expansion in $H_{T}$. In the lowest-order perturbation theory, the electric current $I(V,\Delta T)$
at voltage bias $V$ and temperature gradient $\Delta T$ can be expressed as
\cite{Csonka2012May,Tulewicz2021Jul}
\begin{eqnarray}
I(V,\Delta T)&=&-\frac{2e\Gamma_R}{\hbar} \!\int_{-\infty}^{\infty}\! d\omega A_{L}(\omega) \times \nonumber\\
&&\left[f_L(\omega)-f_R(\omega-eV)\right],
\label{eq:Transport}
\end{eqnarray}
where $A_{L}(\omega)$ is the local density of states (spectral function) of the left subsystem and
$f_\alpha(\omega)$ denotes the Fermi-Dirac distribution function of the lead $\alpha$, 
 ${f_{\alpha}(\omega)= {[1 + \textrm{exp}(\omega/ T_{\alpha})]^{-1}}}$,
with $T_L = 0$ and $T_R = \Delta T$, cf. Fig.~\ref{Fig:E_schematic}, and $k_B \equiv 1$.
\new{The factor of $2$ in \eq{eq:Transport} results from the spin degrees of freedom.}

The spectral function $A_L(\omega)$ of the left subsystem is calculated
by using the numerical renormalization group method \cite{Wilson1975Oct,Bulla2008Apr,FlexibleDMNRG}.
This method is well-suited to account for electron correlations in a very accurate manner,
and especially those giving rise to the Kondo effect \cite{Hewson_1993}.
The spectral function can be related to the imaginary part of the molecule's orbital level
retarded Green's function $G^r_\sigma(\omega)$, 
$A_L(\omega)=\sum_\sigma A_{L\sigma}(\omega)$,
with $A_{L\sigma}(\omega)=-{\rm Im} G^r_\sigma(\omega)/\pi$,
where $G^r_\sigma(\omega)$ is the Fourier transform of $G^r_\sigma(t) = -i\Theta(t)\expect{\{d_\sigma(t),d_\sigma^\dag(0)\}}$.
Within the NRG, we first determine the eigenspectrum of $H_L$
and then calculate $A_L(\omega)$ in the Lehmann representation.
In addition, to improve the quality of the spectral data,
we also employ the z-averaging approach~\cite{Oliveira1990}.

\subsubsection{Seebeck coefficient}

A special emphasis in this paper is put on the nonequilibrium
behavior triggered by a large temperature and/or voltage gradient.
The Seebeck coefficient (or thermopower)
is the thermoelectric property that quantifies the voltage induced by a thermal
gradient across a conductor, and is defined as,
\be
S=-\left(\frac{V}{\Delta T}\right)_{\!\! I=0},
\ee
under the assumption that the current $I$ through the system vanishes, $I=0$.

In the linear response regime, the Seebeck coefficient, $S_{\rm lin}$,
can be reliably described by using the Onsager integrals,
${L_{n} = -\frac{1}{h} \int d\omega \;\omega^n f'(\omega) \mathcal{T}(\omega)}$,
involving the transmission coefficient,
{$\mathcal{T}(\omega) \propto \Gamma_R \,A_L(\omega)$},
where $f'(\omega)$ is the derivative of the Fermi function \cite{barnard1972thermoelectricity}
\be
S_{\rm lin}=-\frac{1}{eT}\frac{L_1}{L_0}.
\label{Eq:Slin}
\ee
This basic definition of the Seebeck coefficient can be directly
extended to the nonequilibrium case by considering
that only the current generated by the thermal gradient must vanish.
Assume, for example, that a system with potential bias $V$ exhibits a current $I$ flowing through it.
When an additional temperature gradient $\Delta T$ is applied,
a new current $I_{\rm tot} = I + I_{\rm th}$ will flow through the system,
where $I_{\rm th}$ is the additional current induced by the thermal gradient.
One can then define a nonequilibrium (nonlinear) Seebeck coefficient $\Sn(V,\Delta T)$ as
\cite{Krawiec2007Apr,Leijnse2010Jul,Azema2014Nov,Jiang2017Jun,
	anderson_LocalizedMagnetic_1961,Jiang2017Jun,PerezDaroca2018Apr,Eckern2020Jan}
\be
\Sn(V,\Delta T) = -\left(\frac{\Delta V}{\Delta T}\right)_{\!\! I(V+\Delta V,\Delta T)=I(V,0)},
\label{Eq:Sn}
\ee
where $\Delta V$ is the change in potential bias required to
suppress the current induced by the thermal gradient $\Delta T$.

Additionally, in the nonlinear response regime,
one can also define a differential Seebeck coefficient
$\Sd(V,\Delta T)$ as \cite{Dorda2016Dec}
\be 
\Sd(V,\Delta T) = -\!\left(\frac{dV}{d \Delta T}\!\right)_{\!\!I} \!=-\!\left(\frac{\partial I}{\partial \Delta T}\!\right)_{\!\! V} \bigg/\!\! \left(\frac{\partial I}{\partial V}\!\right)_{\!\!\Delta T},
\label{Eq:Sd}
\ee
where $\left( \frac{\partial X}{\partial Y} \right)_{\!Z}$ describes the partial derivative of $X$ with respect to $Y$, while keeping $Z$ constant. The differential Seebeck effect is related to the ratio of the thermal response at finite voltage to differential conductance at finite temperature gradient. We note that in the linear response regime with respect to the bias voltage, $\Sd$ becomes comparable to the linear response Seebeck coefficient given by Eq.~(\ref{Eq:Slin}).

\new{
\subsubsection{Heat current and thermoelectric efficiency}

We are also interested in the behavior of the nonequilibrium heat current and
the thermoelectric efficiency $\eta$ \cite{Hershfield2013Aug,Yamamoto2015Oct,Sierra2015Jan,
	Karbaschi2016Sep,Gomez-Silva2018Feb,Yang2020Dec}.
The formula for the heat current can be 
derived from the first law of thermodynamics, which for subsystem $\alpha=L,R$
reads, $dU_\alpha = dW_\alpha + dQ_\alpha$. Here, $dU_\alpha$ is the energy
flowing into the subsystem $\alpha$, while $dQ_\alpha$ denotes the corresponding heat.
The work done to the subsystem $\alpha$ is generally given by,
$dW_\alpha = \mu_\alpha dN_\alpha$, where $dN_\alpha$ is
the corresponding particle number change.
The heat current associated with the left and right subsystem
can be then defined as \cite{Yamamoto2015Oct}
\begin{eqnarray}
I_L^Q(V,\Delta T) &=& I^E(V,\Delta T) - \mu_L I(V,\Delta T) /e, \\
I_R^Q(V,\Delta T) &=& I^E(V,\Delta T) - \mu_R I(V,\Delta T) /e,
\end{eqnarray}
where $I^E(V,\Delta T)$ denotes the energy current given by
\begin{eqnarray}
	I^E(V,\Delta T)&=&-\frac{2\Gamma_R}{\hbar} \!\int_{-\infty}^{\infty}\! d\omega \;\omega\; A_{L}(\omega) \times \nonumber\\
	&&\left[f_L(\omega)-f_R(\omega-eV)\right].
	\label{eq:Transport}
\end{eqnarray}
In our setup, $T_R>T_L$, see \fig{Fig:E_schematic},
such that the electrons flow from the hot reservoir
to the cold one performing the work $\dot{W}(V,\Delta T)$ per unit time.
Such power output can be related to the heat currents through
\begin{equation}
	P\equiv \dot{W}(V,\Delta T) = I_R^Q(V,\Delta T) - I_L^Q(V,\Delta T).
\end{equation}
Then, the thermoelectric efficiency of such a heat engine can be defined as
the ratio of the power to the heat extracted from the hot reservoir
\begin{equation}
	\eta(V,\Delta T) = \frac{\dot{W}(V,\Delta T)} {I_R^Q(V,\Delta T)} = 1-
	\frac{I_L^Q(V,\Delta T)} {I_R^Q(V,\Delta T)}.
	\label{Eq:eta}
\end{equation}
}

\section{\label{sec:results}Results and discussion}

In this section we present and discuss the numerical results
obtained for the differential conductance and Seebeck effect
in far-from-equilibrium conditions.
For the considered system we assume the Coulomb correlations,
$U=0.2$, the coupling to the left lead, $\Gamma_L = 0.02$, in units of half bandwidth,
while the weak coupling to the right lead is assumed to be, $\Gamma_R=\Gamma_L/10$.
In NRG calculations we keep at least $4^5$ states in the iterative procedure
and perform averaging over $8$ different discretizations \cite{Oliveira1990}.

We note that in experimentally relevant scenarios,
a voltage drop imposed across the junction would result in a change
of the orbital level, depending on capacitive couplings to the contacts and to the gate voltage.
However, to get a better understanding of the transport properties,
we first assume that the position of the orbital level does not change as the bias voltage is tuned.
This would correspond to immediately counterbalancing the voltage drop
on the molecule by an appropriate tuning of the gate voltage.
Nevertheless, further on, assuming exemplary parameters of the junctions,
we also present the results for the case when the
orbital level depends on the transport voltage.

\subsection{Electronic transport under finite potential and temperature gradient}

\begin{figure}[t!]
	\centering
\includegraphics[width=0.99\columnwidth]{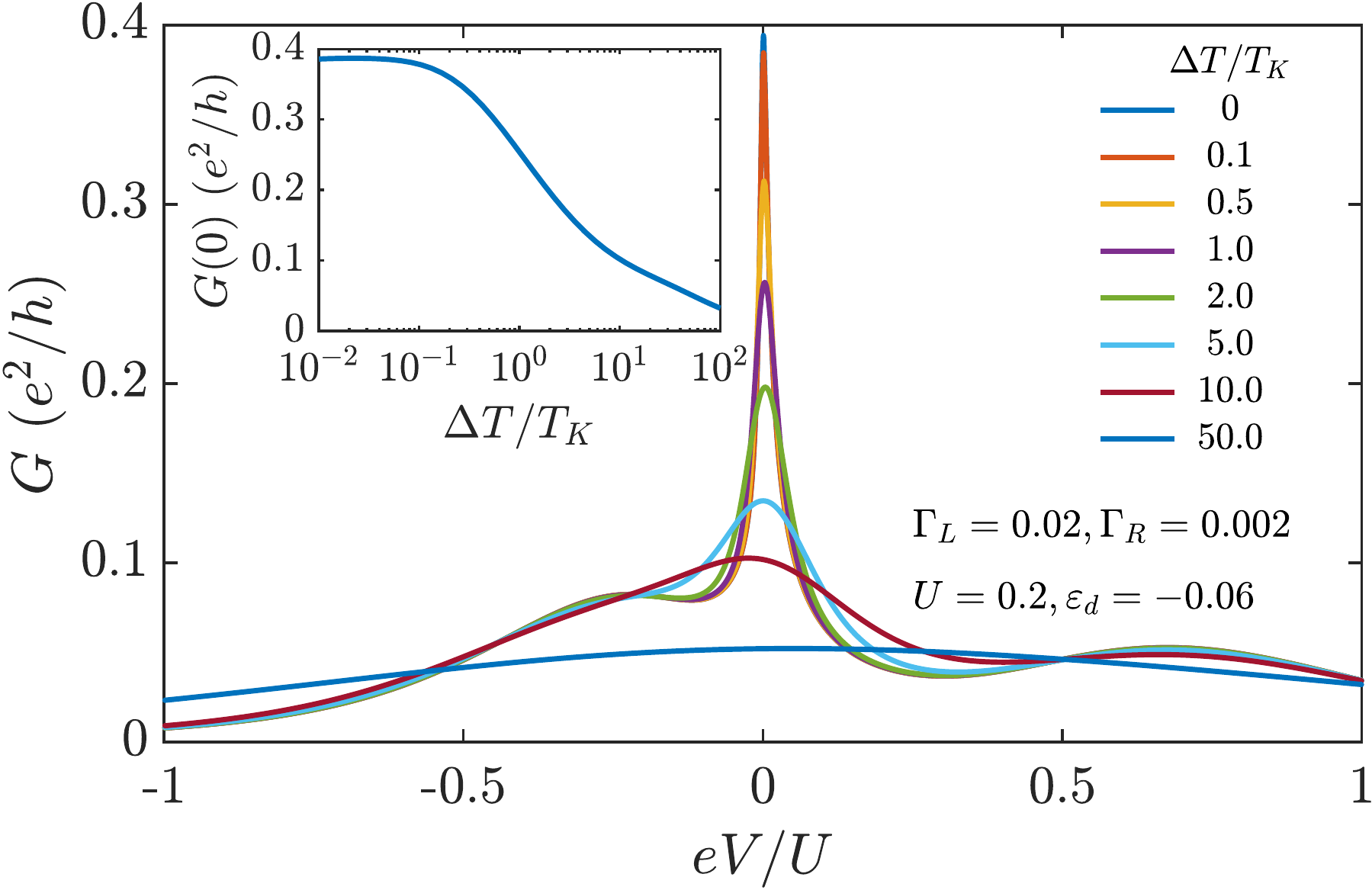}
\caption{The differential conductance plotted as a function of the bias voltage $V$
	for different values of the temperature gradient $\Delta T$, as indicated.
	The inset shows the evolution of the zero-bias conductance as a function of $\Delta T$.
	The parameters are: $U=0.2$, $\Gamma_L=0.02$, $\Gamma_R=0.002$, $T=0$,
	in units of band halfwidth, and $\e_d=-0.3\, U$.
	The Kondo temperature can be estimated from the Haldane's formula,
	which for assumed parameters yields $T_K/U \approx 8.26\cdot 10^{-3}$.
	The Kondo energy scale in the applied bias potential,
	defined as the half-width at half-maximum of the $G(V)$ curve,
	is found to be $eV_K/U \approx 1.45\cdot 10^{-2}$.
}
\label{Fig:I_th}
\end{figure}

\begin{figure*}[t]
	\centering
	\includegraphics[width=0.75\textwidth]{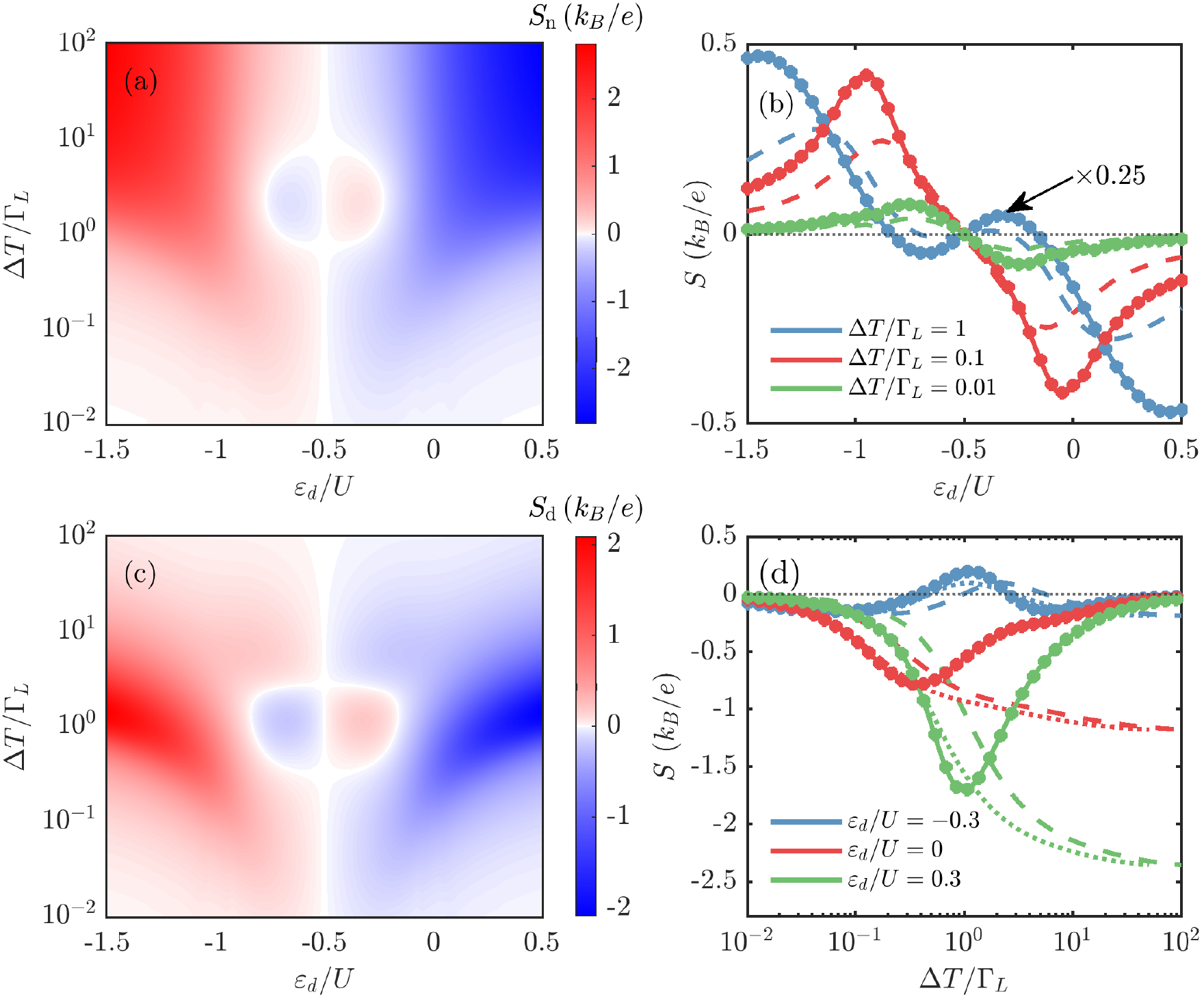}
	\caption{ (a) The non-equilibrium Seebeck coefficient $\Sn$ and
		(c) the differential Seebeck coefficient $\Sd$ calculated as a function of the orbital energy $\e_{d}$
		and the temperature gradient $\Delta T$ at potential bias $V\to 0$
		for parameters the same as in Fig.~\ref{Fig:I_th}.
		Note the logarithmic scale in temperature gradient.
		(b) shows the horizontal cross-sections of (a) and (c),
		i.e., the dependence of $\Sn$ (dashed) and $\Sd$ (solid) as a function of $\e_{d}$
		for various $\Delta T$, whereas (d) presents the vertical cross-sections
		of (a) and (c), i.e.,  the dependence of $\Sn$ (dashed) and $\Sd$ (solid)
		on $\Delta T$ for various $\e_{d}$.
		The colored circles show the linear response thermopower
		$S_{\rm lin} (T=\Delta T)$ and the colored dotted lines in (d) show
		the rescaled non-equilibrium thermopower $\Sn (2 \, \Delta T)$.
	}
	\label{Fig:S}
\end{figure*}

Let us first analyze the behavior of the differential conductance
\new{
$G \equiv dI(V,\Delta T)/dV$
}
as a function of the bias voltage calculated for different temperature gradients $\Delta T$. 
This dependence is presented in \fig{Fig:I_th}. The figure was generated
for $\e_d =-0.3\,U$, i.e. in the local moment regime, where
the system exhibits the Kondo effect 
caused by the strong coupling to the left contact.
The relevant Kondo temperature for the left subsystem
can be estimated from the Haldane's formula \cite{Haldane1978},
which yields, $T_K \approx 8.26\cdot 10^{-3}\,U = 8.26\cdot 10^{-2}\,\G_L $.
Note that because $\Gamma_R = \Gamma_L/10$,
the Kondo temperature associated with the right contact
is exponentially suppressed and thus negligible.
The presence of Kondo correlations 
is reflected in a pronounced zero-bias peak
visible in the differential conductance, see \fig{Fig:I_th}.
Note that because of asymmetric couplings,
the maximum of conductance at zero bias is much reduced
compared to $2e^2/h$.
With increasing the bias voltage, the conductance decreases
and shows smaller resonances corresponding to $eV\approx \e_d$
and $eV\approx \e_d+U$. When the thermal gradient increases
(note that the base system temperature is assumed to be $T=0$),
one observes a gradual suppression of the zero-bias anomaly,
until the whole bias dependence of $G$ does not show any
Kondo correlation effects for $\Delta T \gg T_K$. 
The evolution of the Kondo peak with increasing $\Delta T$
is explicitly presented in the inset of \fig{Fig:I_th}.
The conductance drops to a half of its maximum value
when $\Delta T \approx 2 T_K$. This reflects the fact
that the actual system temperature, which can be 
associated with an average of the left and right contact
temperatures, is equal to $\Delta T/2$.
We also note that for $\Delta T \ll T_K$
one can quantify the Kondo resonance by $V_K$,
which characterizes the Kondo energy scale in the applied bias potential,
defined as the half-width at half-maximum of the $G(V)$ curve.
For $V_K$ we find, $eV_K \approx 1.45\cdot 10^{-2} \, U = 1.45\cdot 10^{-1} \, \G_L$.

\subsection{Non-equilibrium thermopower}

In this section we focus on the analysis of the 
behavior of thermopower, both $\Sn$ and $\Sd$, cf. Eqs.~(\ref{Eq:Sn}) and (\ref{Eq:Sd}),
under finite temperature and voltage gradients.
However, to get a better understanding of thermoelectric transport,
we first start with the case of $V\to 0$,
while nonequiliibrium settings are imposed only by increasing $\Delta T$.
The more general case of having both finite $\Delta T$ and $V$ will be examined afterwards.

\subsubsection{Thermopower under finite temperature gradient}

The nonequilibrium $(\Sn)$ and differential $(\Sd)$ Seebeck coefficients
as a function of temperature gradient
$\Delta T$ and orbital level position $\varepsilon_d$,
for the case when the nonlinear response regime is triggered by a large
temperature gradient, are presented in \fig{Fig:S}.
Before analyzing the behavior of the thermopower in greater detail,
let us first briefly discuss different regimes
for the energy of the orbital level $\e_d$, and what it implies.
The local moment regime, $-1\lesssim \e_d/U \lesssim 0$, denotes the value of orbital energy,
in which the singly occupied level is held below the Fermi energy
(of the left electrode in our case) and the doubly occupied state
is above the Fermi level. This is the regime where the molecule
is occupied by an unpaired electron and the system can exhibit the Kondo effect.
As can be inferred from the name, the empty/fully occupied regime,
$\e_d/U\gtrsim 0$, $\e_d/U\lesssim -1$,
refers to the case where the preferred configuration is having the orbital
level completely empty or fully occupied. On the other hand,
when $\e_d/U\approx-1$ or $\e_d/U\approx 0$,
we reach a mixed valence/resonant tunneling regime where the orbital level
is in the vicinity of the Fermi level of the electrode (depending on its hybridization $\Gamma_L$).

\begin{figure}[t]
	\centering
	\includegraphics[width=0.9\columnwidth]{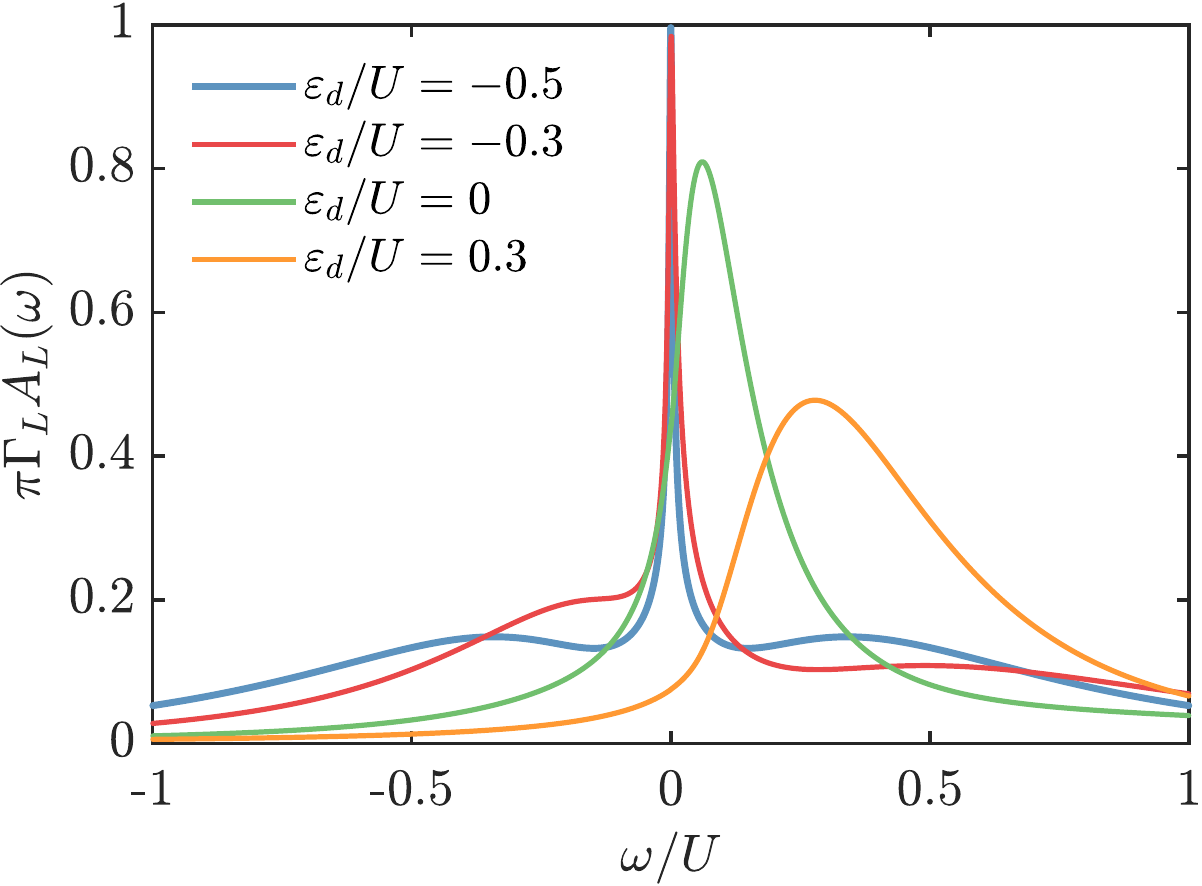}
	\caption{The normalized spectral function $\pi\Gamma_L A_L(\omega)$
		of the left subsystem calculated for various orbital energy $\e_d$, as indicated.
		The other parameters are the same as in \fig{Fig:I_th}.
	}
	\label{Fig:A}
\end{figure}

The colormaps for $\Sn$ and $\Sd$ as presented in \figs{Fig:S}(a) and (c)
show a similar behavior but with interesting deviations.
We note that for $V\to 0$ both thermopowers are odd functions of $\e_d$
across the particle-hole symmetry point,
$\e_d=-U/2$ [see also \fig{Fig:S}(b)],
and decay to zero when the temperature gradient $\Delta T \rightarrow 0$
[see also \fig{Fig:S}(d)].
Moreover, both $\Sn$ and $\Sd$ are generally negative (positive) for $\e_d>-U/2$ ($\e_d<-U/2$),
indicating a dominant role of the electron (hole) processes.
When inside the local moment regime,
both $\Sn$ and $\Sd$ survive to even lower values of temperature gradient $\Delta T/\Gamma_L \approx 10^{-2} $,
compared to $\Delta T/\Gamma_L \approx -10^{-1}$, as in the case of empty/fully occupied regime.
This is due to the presence of Kondo correlations when the orbital level is singly occupied.
Additionally, in this transport regime
both thermopowers change sign twice as a function of $\Delta T$.
The first sign change occurs when $1 \lesssim \Delta T/\Gamma_L\lesssim 10$,
whereas the second one develops when 
$1/10 \lesssim \Delta T/\Gamma_L\lesssim 1$, see \figs{Fig:S}(a) and (c).
The sign changes of thermopower in the linear response can be 
assigned to the corresponding behavior of the transmission coefficient
(in our case the spectral function of the left subsystem) \cite{Costi2010Jun,Weymann2013Aug}.
The Sommerfeld expansion indicates that it is the slope of 
$\mathcal{T}(\omega)\propto A_L(\omega)$ at the Fermi level,
which determines the sign of the Seebeck coefficient \cite{Costi2010Jun}.
Of course, this strictly holds in the low-temperature regime,
however, it also allows for shedding some light onto the nonequilibrium behavior
where in turn the dependence of $S$ is determined by the whole integral
in Eq.~(\ref{eq:Transport}).
Therefore, for the sake of completeness, in
\fig{Fig:A} we present the energy dependence of the normalized
spectral function of the left subsystem calculated for different values
of the orbital level position, as indicated.
In the local moment regime, a pronounced Kondo peak can be seen
at the Fermi energy accompanied by two Hubbard resonances
at $\omega = \e_d$ and $\omega = \e_d+U$.
On the other hand, in the mixed valence regime
a resonant peak with position around $\omega=0$, renormalized
by the coupling strength, develops, which moves to positive energies
when raising the orbital level, see the case of $\e_d/U=0.3$ in \fig{Fig:A}.
Having that in mind, one can qualitatively understand the sign changes in the local moment regime.
When lowering the temperature, the first sign change
corresponds to change of slope of $A_L(\omega)$ around the Hubbard peak,
whereas the second sign change has been found to indicate the onset of the Kondo effect.
Note, however, that the value of temperature gradient corresponding
to the sign change does not correspond to the Kondo temperature
of the system but just indicates the emergence of Kondo correlations \cite{Costi2010Jun}. 

The differences between $\Sn$ and $\Sd$ become evident
for higher values of the temperature gradient $\Delta T$.
Specifically, beyond the local moment regime,
$\Sn$ starts to saturate above $\Delta T/\G_L \gtrsim 1$,
whereas $\Sd$ reaches maximum around 
$\Delta T/\G_L \approx 1$ and then becomes suppressed
with increasing $\Delta T$, decaying to zero for $\Delta T/\G_L \gtrsim 10$.
This can be explicitly seen in \fig{Fig:S}(d),
which presents the vertical cross-sections of \figs{Fig:S}(a) and (c).
The difference in the high temperature behavior of $\Sn$ and $\Sd$
highlights the fundamental difference in the definitions of both thermopowers.
The differential Seebeck coefficient $\Sd$
corresponds essentially to a response function and can justifiably
show a vanishing response when the temperature gradient gets high enough.
On the other hand, the nonequilibrium Seebeck coefficient $\Sn$ indicates
the magnitude of the potential difference developed across
the junction with a finite $\Delta T$, which even at high temperature gradient has to remain finite.
There is also a noticeable difference in the '$\Phi$'-like shape
visible in \figs{Fig:S}(a) and (c),
drawn by the points of sign-change for each thermopower,
which originates from the differences in the $\Delta T$ behavior
of both $\Sn$ and $\Sd$ and can be understood from
the below discussion of the cross-sections of the colormaps.

\begin{figure*}[t]
	\centering
	\includegraphics[width=0.75\textwidth]{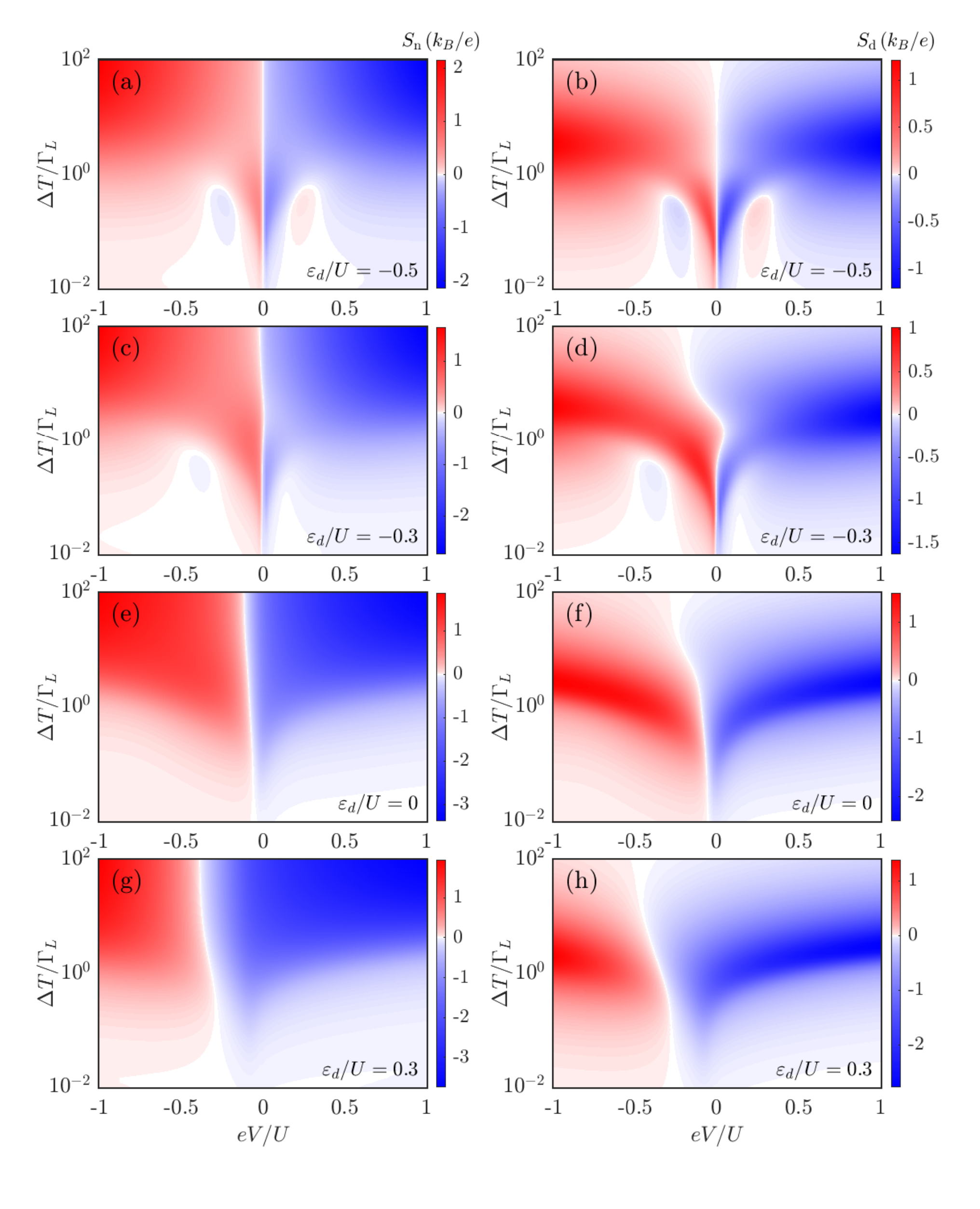}
	\caption{(Left column) The nonequilbrium Seebeck coefficient $\Sn$
		and (right column) the differential Seebeck coefficient $\Sd$
		calculated as a function of the temperature gradient $\Delta T$
		and the applied bias voltage $V$.
		(a) and (b) present the case of the particle-hole symmetry point $\varepsilon_d/U = -0.5$,
		(c) and (d) show the local moment regime with $\varepsilon_d/U = -0.3$,
		(e) and (f) display the resonant tunneling regime $\varepsilon_d/U = 0$,
		whereas (g) and (h) present the empty orbital regime with $\varepsilon_d/U = 0.3$.
	}
	\label{Fig:S_V}
\end{figure*}

\begin{figure*}[t]
	\centering
	\includegraphics[width=0.75\textwidth]{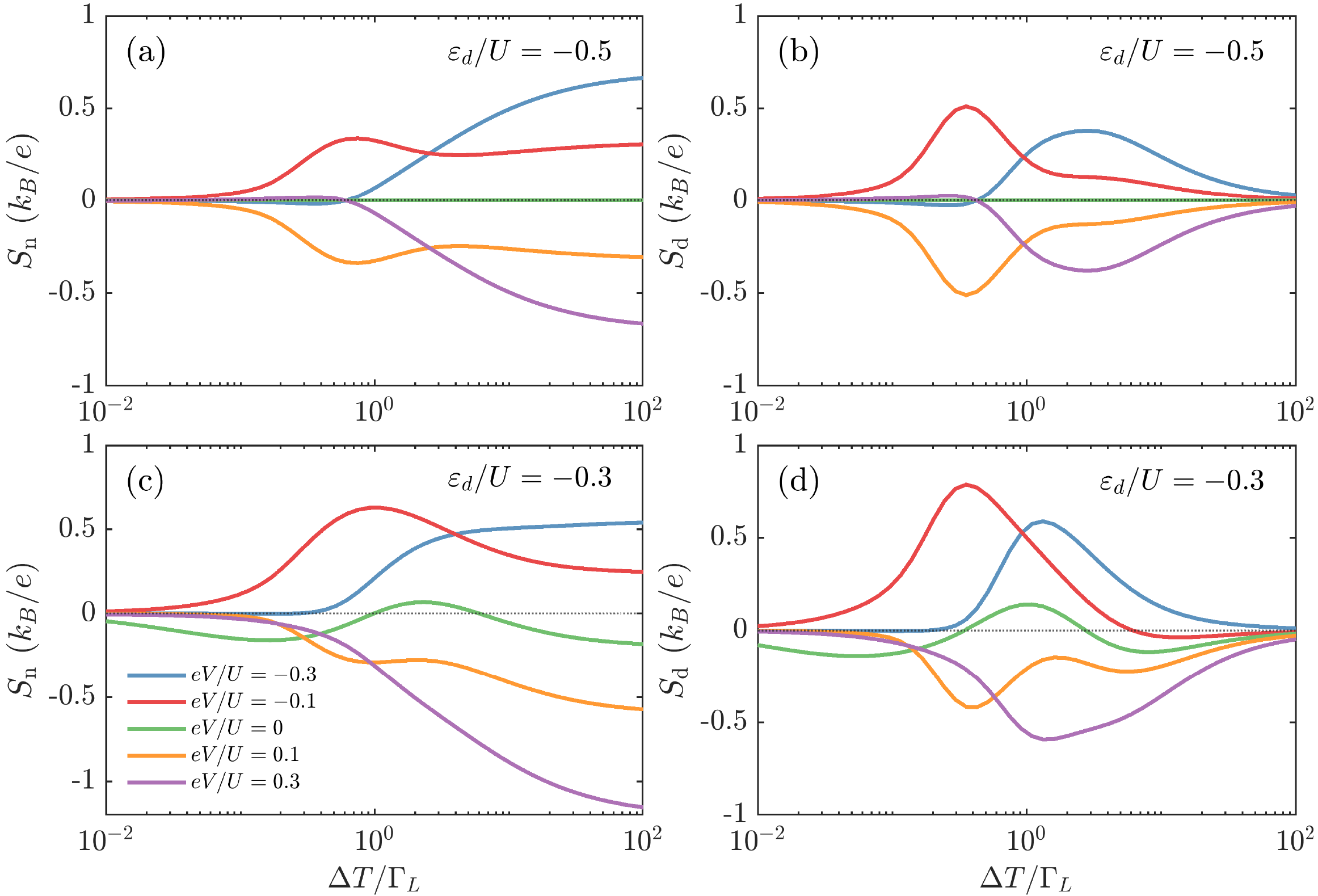}
	\caption{The temperature gradient dependence of (left column)
		the nonequilibrium and (right column) the differential Seebeck coefficients for different values of the 
		bias voltage applied to the system, as indicated.
		(a) and (b) present the case of $\varepsilon_d/U=-0.5$,
		while (c) and (d) display the case of  $\varepsilon_d/U=-0.3$, respectively.
		This figure corresponds to the vertical cross-sections of Figs.~\ref{Fig:S_V}(a)-(d).
	}
	\label{Fig:Sn_dT2}
\end{figure*}

The fact that $\Sn$ and $\Sd$ for $V \rightarrow 0$ and $\Delta T \rightarrow 0$
should recover the linear response thermopower $S_{\rm lin}$
justifies the comparison of $\Sn(\Delta T)$ and $\Sd(\Delta T)$ to $S_{\rm lin}(T)$.
The cross-sections in \figs{Fig:S}(b) and (d)
compare $\Sn$ and $\Sd$ as a function of $\e_d$ and $\Delta T$, respectively,
along with the linear response thermopower $S_{\rm lin}$
calculated for the system with global temperature $T= \Delta T$, denoted as colored circles. 

We note that the dependence of the differential Seebeck coefficient $\Sd$
over the temperature gradient $\Delta T$ shows good agreement
with the behavior of the linear response Seebeck effect $S_{\rm lin}$
with the global temperature $T$ even at large temperature gradients.
This is not surprising due to the fact that for $\mu_{\rm L}=0$ and $T_{\rm L}=0$,
Eq.~(\ref{eq:Transport}) gives a formally similar expression
for $S_{\rm lin}(T)$ and $\Sd(\Delta T)$. 
It is also interesting to note that, although $\Sn$ deviates
from this behavior at large temperatures, the $\Sn(\Delta T)$ dependence
when rescaled by 2 in $\Delta T$, i.e. $\Sn(2 \Delta T)$
[see the dashed lines in \fig{Fig:S}(d)],
agrees well with the linear response behavior for low values of $\Delta T$, see \fig{Fig:S}(d).

To further understand the behavior of the Seebeck coefficients in \fig{Fig:S},
one can separately focus on the three regimes.
Since for $V\to 0$ thermopower is an odd function across the particle-hole symmetry point,
we pick representative values of $\e_d$ for $\e_d>-U/2$.
In the local moment regime, see the case of $\varepsilon_d/U = -0.3$ in \fig{Fig:S}(d),
$\Sn$ first starts decreasing with raising $\Delta T$
until  $\Delta T \approx T_K$ and then it increases to reach a sign change
occurring around $\Delta T/\G_L \approx 1$.
Further increase of $\Delta T$ however gives rise to another sign change
and the thermopower becomes again positive for large temperature gradients.
Such behavior is consistent with the linear-response considerations
of Kondo-correlated systems \cite{Costi2010Jun,Weymann2013Aug}.
Here, however, we demonstrate that it extends to nonlinear response regime.
The above-described behavior is absent when the orbital level is tuned out of the Kondo regime.
For resonant conditions, i.e. $\varepsilon_d = 0$,
$\Sn$ grows only in a monotonic fashion with increasing $\Delta T$.
The same can be observed for the empty orbital case ($\varepsilon_d /U = 0.3$),
but now the value of $\Sn$ gets enhanced for $\Delta T / \G_L\gtrsim 1$,
compared to the case of $\varepsilon_d = 0$.

\begin{figure*}[t]
	\centering
	\includegraphics[width=1.02\textwidth]{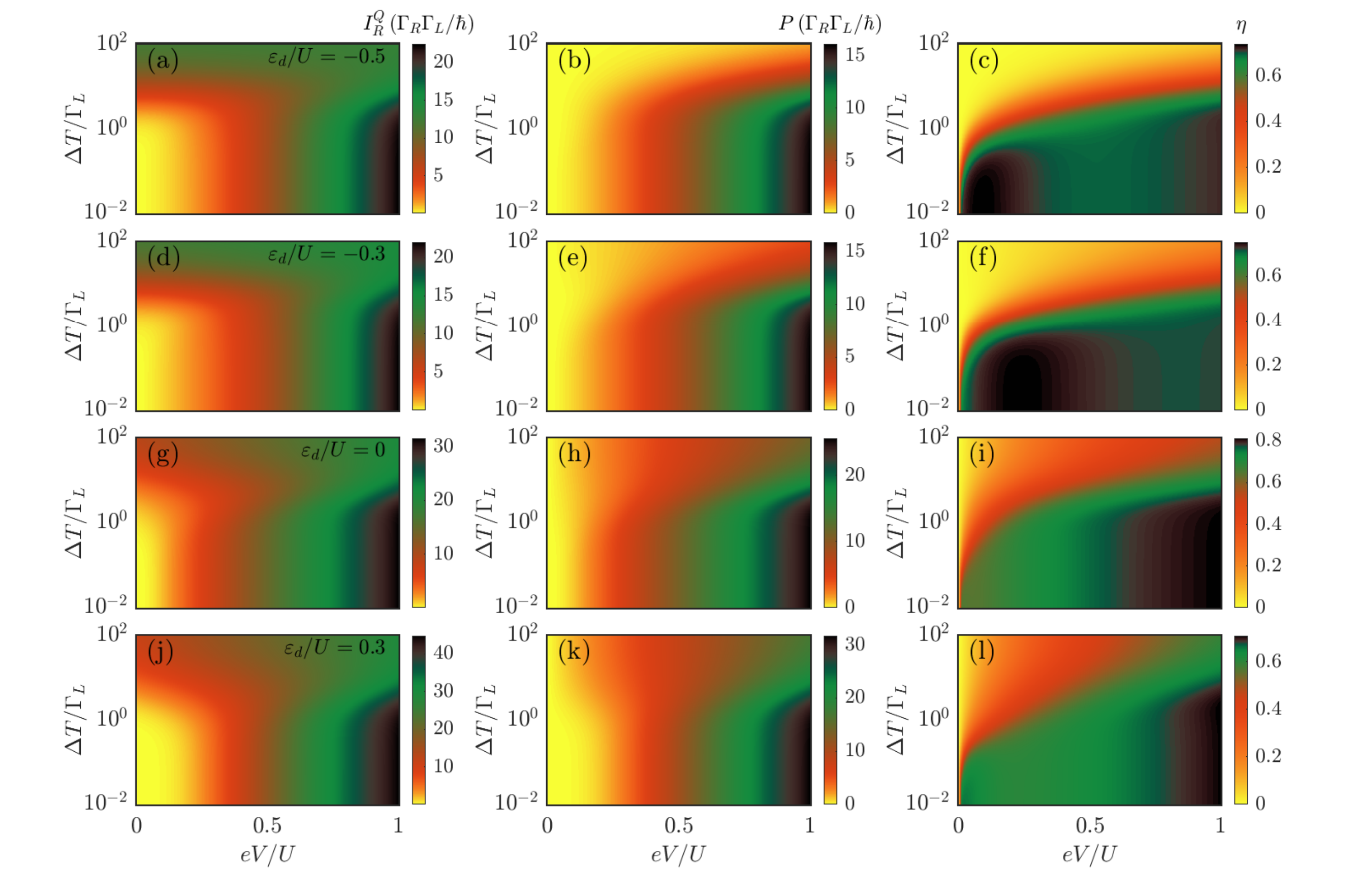}
	\caption{(left column) The heat current $I_R^Q$,
		(middle column) the power $P\equiv I_R^Q-I_L^Q$,
		as well as (right column) the thermoelectric efficiency $\eta$
		calculated as a function of the applied bias voltage $V$
		and the temperature gradient $\Delta T$
		for different values of orbital energy $\varepsilon_d$.
		The first row is determined for $\e_d/U=-0.5$,
		the second one for $\e_d/U=-0.3$,
		the third row is for $\e_d=0$,
		while the last one is calculated for $\e_d/U=0.3$.
		The other parameters are the same as in \fig{Fig:I_th}.
	}
	\label{Fig:eta}
\end{figure*}

\subsubsection{Thermopower for finite temperature gradient and bias voltage}

After analyzing the influence of the nonlinear temperature gradient
on the Seebeck coefficient at linear potential bias,
we will now proceed with the examination of the thermopower
in the presence of both finite bias voltage and temperature gradient.
Figure~\ref{Fig:S_V} presents the nonequilibrium and differential Seebeck coefficients calculated
for finite potential and temperature gradients
for different values of the orbital level energy, as indicated.

Let us first focus on the case when at low voltages the system is in the local moment regime.
This is presented in \figs{Fig:S_V}(a,b) for the particle-hole symmetry point ($\varepsilon_d/U=-0.5$),
and in \figs{Fig:S_V}(c,d) when detuned out of the symmetry point ($\varepsilon_d/U=-0.3$).
From the linear response studies,
one can expect the thermopower at the particle-hole symmetry point
to vanish, since $S$ can be related to the slope of the spectral function
$\frac{\partial A_L(\omega)}{\partial(\omega)}$ around the Fermi energy of the left lead in our case.
In our system, where the potential bias is applied asymmetrically (only on the right lead),
we are essentially shifting the system out of the symmetry point with an applied bias potential.
This results in a non-zero Seebeck effect for a finite potential bias
even in the case of $\varepsilon_d=-U/2$, see the first row of \fig{Fig:S_V}.
As can be seen, in the nonlinear response regime for $\varepsilon_d=-U/2$,
both $\Sn$ and $\Sd$ become finite and are odd functions of the applied bias voltage.
More specifically, we observe positive Seebeck coefficients
for negative applied bias voltage and vice versa.
When the temperature gradient is of the order of the Kondo temperature,
there is a new region of sign change present in the thermopower
when the potential bias is between the range $0.2\lesssim|eV/U|\lesssim0.4$,
i.e. the system is in the Coulomb blockade.
This sign change, visible in both $\Sn$ and $\Sd$ at nonequilibrium settings, is due to the Kondo correlations. 
The sign changes of the differential Seebeck coefficient $\Sd (V)$
at low temperature gradient $\Delta T$ can be inferred
directly from the slope of the spectral function.
From \Eq{Fig:I_th}, when $|eV|\gg\Delta T$, the thermal response $\frac{\partial I}{\partial \Delta T}$
does not change considerably, resulting in the sign of $\Sd$ to be entirely dependent
on $G=\frac{\partial I}{\partial V}$, cf. \Eq{Eq:Sd}.
This mandates the sign of $\Sd$ to roughly follow
the sign of the function $f\new{(eV)} = -(\frac{\partial A_L(\omega = e V)}{\partial \omega})^{-1}$. 
When the bias voltage increases further, 
the Fermi levels of the leads are too far apart to show any effect of Kondo correlations on transport.
In other words, the system leaves the Coulomb blockade regime for $|eV/U|\gtrsim 0.5$.
For the case of the orbital level detuned out of the p-h symmetry point,
we observe that the dependence of thermopower becomes generally asymmetric with
respect to the bias reversal. For  $\varepsilon_d/U=-0.3$,
the sign changes corresponding to the Kondo correlations
are present only for negative bias voltage, where the regime of sign change
moves to slightly more negative bias voltage, $-0.5\lesssim eV/U\lesssim -0.25$.
Interestingly, there is a new sign change, seen mostly in the temperature dependence 
of $\Sd$ for small positive bias voltages, which develops around $\Delta T/\G_L \approx 1$,
see  \fig{Fig:S_V}(d).

All the features discussed in the case when at low voltages the system is
in the local moment regime are nicely exemplified in \fig{Fig:Sn_dT2},
which presents the relevant vertical cross-sections of \fig{Fig:S_V}.
One can clearly see the development of a finite Seebeck effect
with increasing the potential bias in the case of $\e_d/U=-0.5$,
where a small sign change for $|eV|/U=0.3$ can be observed.
Note also the symmetry with respect to the bias reversal.
On the other hand, even larger and generally nonzero for any value of $V$
thermopower develops for $\e_d/U=-0.3$. Here, one can explicitly
see the sign changes of both $\Sn$ and $\Sd$ in the low bias voltage regime,
which then disappear when the bias voltage increases.

The case when at equilibrium the orbital level is outside the local moment regime
is presented in \figs{Fig:S_V}(e,f) for $\e_d=0$ and in \figs{Fig:S_V}(g,h)
for the empty orbital case $\e_d/U=0.3$.
Now, we generally observe only one sign change
visible in the Seebeck coefficients $\Sn$ and $\Sd$ as a function of the applied bias voltage.
Moreover, there is also a sign change as a function of thermal gradient for selected value of $V$.
In the case of $\varepsilon_d/U=0$, the sign change is close to zero voltage
and the offset from $V=0$ can be explained by
renormalization of the orbital level by charge fluctuations with the strongly
coupled lead, which give rise to a resonance in $A_L(\w)$
slightly shifted with respect to the Fermi energy, cf. \Fig{Fig:A}.
When the potential bias $V$ is positive,
both Seebeck coefficients, $\Sn$ and $\Sd$, are found to be negative.
Moreover, in the regions of negative voltages,
the magnitude of voltage required to change the sign of the thermopower
increases with raising the temperature gradient.
We observe a similar behavior in the case of $\varepsilon_d/U=0.3$, 
but the sign change of the Seebeck effect with respect to the potential bias
is offset by the value of $\varepsilon_d$ in the negative voltage direction,
see the last row of \fig{Fig:S_V}.

\new{
\subsection{\label{sec:heat}Nonequilibrium heat current and the thermoelectric efficiency}

Figure~\ref{Fig:eta} presents the bias voltage and temperature gradient
dependence of the heat current $I_R^Q$ associated with the right subsystem,
the power $P$ generated by the device
together with the thermoelectric efficiency, cf. \eq{Eq:eta},
calculated for several values of the dot level position, as indicated.
First of all, we note the general tendency 
to increase the heat current by raising $V$ or $\Delta T$, 
which is irrespective of the level position $\e_d$.
Moreover, for large bias voltages,
there is a saturation and a slight decrease of $I_R^Q$
with increasing $\Delta T$.
The same can be observed for the power generated by this device.
Up to $eV\sim U/2$, the increase of the temperature gradient
gives rise to an enhancement of the power. However, 
for larger voltages, there is a nonmonotonic dependence of $P$ with respect to $\Delta T$.
The efficiency of the system is presented in the right column of \ref{Fig:eta}.
One can clearly identify an optimal choice of both $\Delta T$ and $V$,
for which $\eta$ becomes maximized.
The parameter space with maximum $\eta$
strongly depends on the transport regime.
Interestingly, in the local moment regime, 
see Figs.~\ref{Fig:eta}(c) and (f), the maximum efficiency
is obtained just around the Kondo regime.
On the other hand, out of the Coulomb blockade and Kondo regime,
the maximum efficiency occurs for larger voltages $eV/U\gtrsim 0.5$,
see the case of $\e_d=0$ and $\e_d/U=0.3$ in 
Figs.~\ref{Fig:eta}(i) and (l).
We also note that the Carnot efficiency for our system where $T_L\to 0$ is $\eta_C=1$.
Consequently, we predict that the efficiency of the considered device
can reach up to $\eta/\eta_C \approx  0.8 $, depending on the transport regime.
}

\subsection{\label{sec:real} Realistic junction}

\begin{figure}[bt]
	\centering
	\includegraphics[width=0.48\columnwidth]{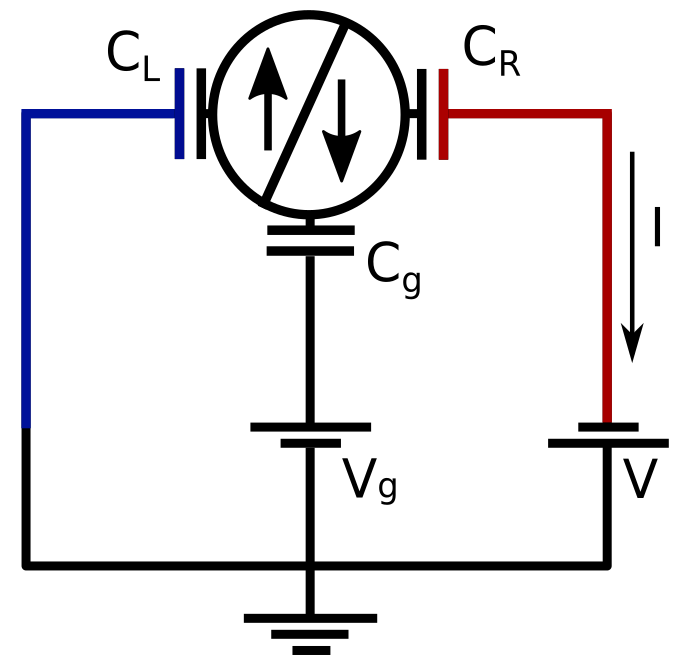}
	\caption{The schematic shows an equivalent electrical circuit diagram
		for the asymmetrically coupled molecular junction
		along with the implementation of the bias potential $V$
		and the gate potential $V_g$. The capacitances associated with the left, right
		and the gate electrodes are represented as $C_L$, $C_R$ and $C_g$, respectively.
	}
	\label{Fig:Real_model}
\end{figure}

\begin{figure*}[t]
	\centering
	\includegraphics[width=0.7\textwidth]{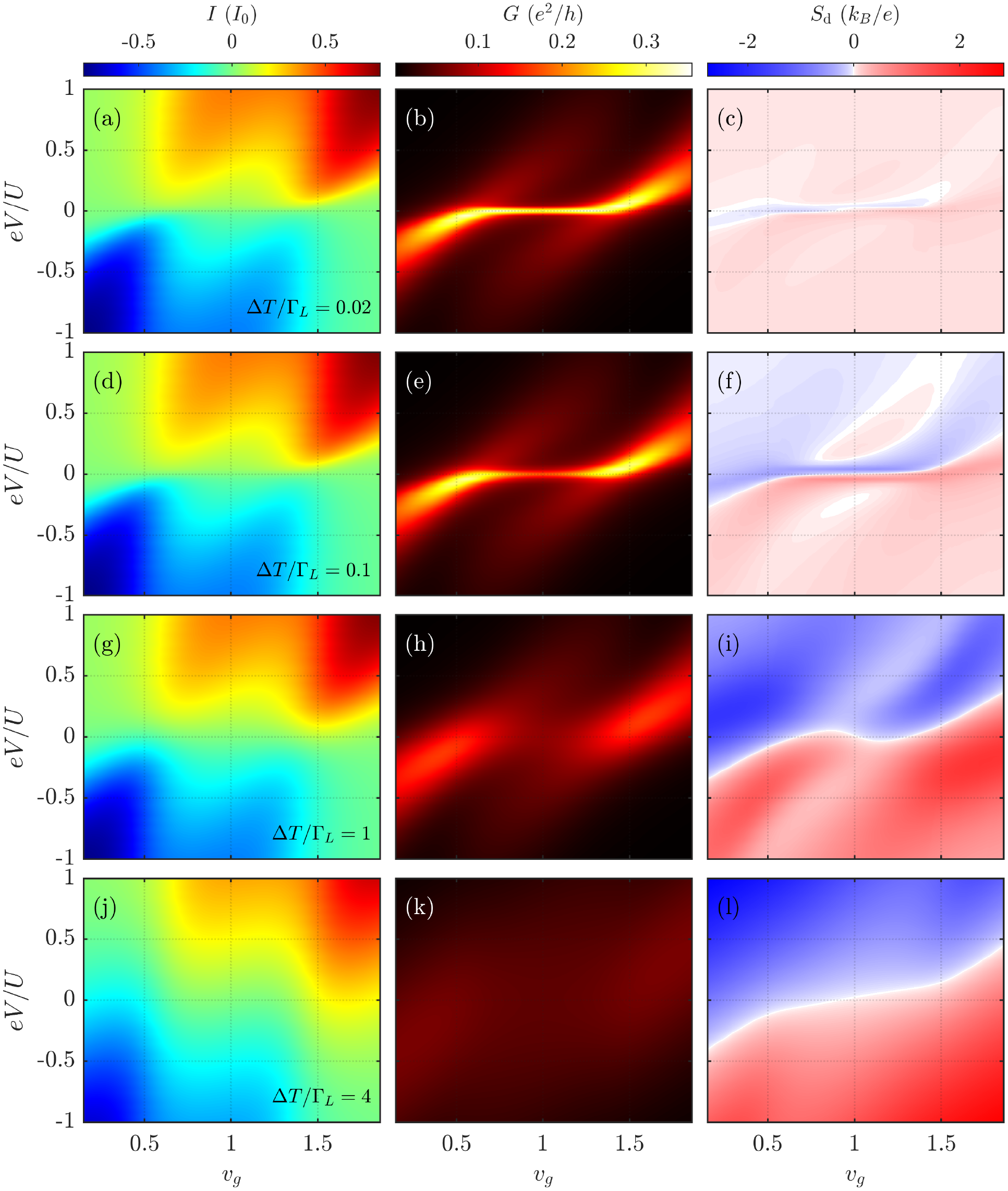}
	\caption{(First column) The current $I$ in units of $I_0=2e\G_R/\hbar$,
		(second column) the differential conductance $G$
		and (third column) the differential Seebeck coefficient $\Sd$ calculated
		as a function of the applied bias voltage $V$ and
		the effective gate voltage $v_g$.
		For the capacitances we assume
		$C_L/C_R=2$ and $C_g/C_R = 0.1$
		and the other parameters are the same as in \fig{Fig:I_th}.
		Each row corresponds to different temperature gradient $\Delta T$, as indicated.
	}
	\label{Fig:Real}
\end{figure*}

To make the discussion of nonequilibrium thermopower more appealing to experimental realizations,
in this section we relax the condition of voltage-independent orbital level
and include the voltage drops assuming realistic junction parameters.
An electrical circuit diagram of the considered system is shown in \fig{Fig:Real_model}.
The tunnel junctions are characterized by the capacitances 
$C_L$ and $C_R$, and the gate capacitance is denoted by $C_g$ with gate voltage $V_g$.
The formula for the current flowing through such system can be written as
	\begin{eqnarray}
	I(V,\Delta T)&=&-\frac{2e\Gamma_R}{\hbar} \!\int_{-\infty}^{\infty}\! d\omega A_{L}(\omega,v_g-v) \times \nonumber\\
	&&\left[f_L(\omega)-f_R(\omega-eV)\right],
	\label{eq:Transport2}
\end{eqnarray}
where $v_g = C_g V_g/e$ and $v=V C_R /e$ are the dimensionless
gate and bias voltage drops  \cite{Csonka2012May}.
For the junction capacitances, we assume $C_L/C_R=2$ and $C_g/C_R = 0.1$,
while the charging energy $E_C = e^2/2C$, with $C=C_L+C_R+C_g$, is equal to $E_C=U/2$.

The current, differential conductance and differential Seebeck effect
as a function of the bias voltage $V$ and the dimensionless gate voltage $v_g$
calculated for different temperature gradients are shown in \fig{Fig:Real}.
At low bias voltages, $v_g$ sets the corresponding transport regimes:
$v_g\lesssim 0.5$ ($v_g\gtrsim 1.5$) defines the empty (fully occupied)
orbital regime, whereas the local moment regime is realized for 
$0.5 \lesssim v_g \lesssim 1.5$. 
Let us first analyze the case of the lowest temperature gradient,
shown in the first row of \fig{Fig:Real}, which corresponds
to the situation when $\Delta T < T_K$.
In \fig{Fig:Real}(a) one can see a clear Coulomb diamond structure.
On the other hand, the differential conductance exhibits
then a pronounced zero-bias Kondo peak in the singly occupied orbital regime,
i.e. $0.5 \lesssim v_g \lesssim 1.5$. Note, that the behavior 
of the differential conductance is typical for tunnel junctions
with asymmetric couplings to the contacts. Alternatively, one could
also think of an adatom probed by a scanning tunneling microscope tip,
which would correspond to a weakly coupled lead.
At low temperatures, the differential thermopower is generally relatively small,
however, one can still recognize sign change at low bias voltage regime,
see \fig{Fig:Real}(c).

When the thermal gradient becomes comparable to the Kondo temperature,
the thermal response gets enhanced. This is presented in the second row of \fig{Fig:Real}.
Although the current is hardly affected, the differential conductance
shows a suppression of the Kondo resonance at zero bias voltage.
Moreover, the differential Seebeck coefficient is increased 
at low bias voltages. In addition, in the Kondo regime
one can clearly see a sign change of $\Sd$ as a function of positive bias voltage. 

With increasing the temperature gradient even more, see the case of $\Delta T/\G_L = 1$
in \fig{Fig:Real}, the Coulomb diamond structure becomes smeared,
and so does the conductance, which now only shows broad resonances
due to resonant tunneling processes. The differential thermopower, on the other hand,
is now much enlarged and it generally displays two regimes
of either negative or positive values. Note, however, that the
wavy line along which $\Sd$ changes sign strongly depends on both $V$ and $v_g$,
which is due to the voltage dependence of the orbital level.
Finally, for larger temperature gradients, $\Delta T/\G_L > 1$, see the last row of
\fig{Fig:Real}, most of the features are smeared. The differential conductance
is suppressed by the thermal fluctuations, whereas the differential Seebeck effect
again shows two regimes with different signs, but now separated by 
a line that monotonously depends on $v_g$.

\section{\label{sec:summary}Summary}

In this paper we have studied the nonequilibrium thermoelectric transport 
properties of a molecular junction comprising a single-orbital molecule (or a quantum dot)
asymmetrically attached to external electrodes. For such setup,
we have determined the nonequilibrium electric and heat currents flowing through the system
by using the perturbation theory with respect to the weakly coupled contact, while
the strongly coupled subsystem was solved by using the numerical renormalization group method.
This allowed us to accurately take into account
the electronic correlations that result in the development
of the Kondo effect between the molecule and strongly coupled lead.
In particular, we have determined the temperature gradient and
bias voltage dependence of the nonlinear $\Sn$ and differential $\Sd$ Seebeck coefficients.
First, we have performed the calculations assuming that 
the orbital level position is independent of the applied bias.
Then, assuming realistic junction parameters,
we have also considered the case when this condition is relaxed.
In particular, we have shown that both $\Sn$ and $\Sd$
exhibit sign changes at nonequilibrium conditions,
which are due to Kondo correlations.
Up to now, such Kondo-related sign changes have been mostly
observed in the linear response regime \cite{Costi2010Jun}.
\new{
In addition, we have also determined the nonequilibrium heat currents,
the power generated by the device, when it works as a heat engine,
and the corresponding thermoelectric efficiency. We have found
transport regimes characterized by a considerable efficiency of up to $80\%$ of the Carnot efficiency.
}
We believe that our results shed new light on the thermopower
of strongly-correlated molecular junctions in out-of-equilibrium settings
and will foster further efforts in the examination of such systems.


\begin{acknowledgments}
	This work was supported by the Polish National Science
	Centre from funds awarded through the decision Nos.~2017/27/B/ST3/00621 and 2021/41/N/ST3/02098.
	We also acknowledge the computing time at the Pozna\'{n} Supercomputing and Networking Center. 
\end{acknowledgments}

\bibliography{asymmetric_QD}

\end{document}